\documentclass[12pt]{article}
  \usepackage{amsfonts}
  \usepackage{amsmath}
\usepackage{amssymb}
\usepackage{amscd}
\usepackage[dvips]{graphicx}
\begin{document}
~~
\bigskip
\bigskip
\begin{center}
{\Large {\bf{{{Canonical, Lie-algebraic and quadratic twist
deformations of  Galilei group}}}}}
\end{center}
\bigskip
\bigskip
\bigskip
\begin{center}
{{\large ${\rm {Marcin\;Daszkiewicz}}$ }}
\end{center}
\bigskip
\begin{center}
{ {{{Institute of Theoretical Physics\\ University of Wroc{\l}aw pl.
Maxa Borna 9, 50-206 Wroc{\l}aw, Poland\\ e-mail:
marcin@ift.uni.wroc.pl}}}}
\end{center}
\bigskip
\bigskip
\bigskip
\bigskip
\bigskip
\bigskip
\bigskip
\bigskip
\begin{abstract}
New Galilei quantum  groups dual to the Hopf algebras proposed in
\cite{dasz} are obtained by the nonrelativistic contraction
procedures. The corresponding Lie-algebraic and quadratic quantum
space-times are identified with  the translation sectors of
considered algebras.
\end{abstract}
\bigskip
\bigskip
\bigskip
\bigskip
\eject
\section{{{Introduction.}}}

Recently, there were found arguments based on quantum gravity
\cite{grav1}, \cite{grav2} and string theory \cite{string1},
\cite{string2} indicating  that space-time at Planck scale  should
be noncommutative, i.e. it should  have a quantum nature. On the
other side, there appeared a lot of papers dealing with classical
(\cite{mech1}-\cite{mechn}) and quantum
(\cite{qm1}-\cite{lodzianieosc}) mechanics, Doubly Special
Relativity frameworks (\cite{dsr1a}-\cite{dsr1c}), and field
theoretical models (\cite{field1}-\cite{fieldn}), in which
noncommutative
 space-time plays a prominent role.

In accordance with general
 classification of all possible
deformations of relativistic and nonrelativistic symmetries
(\cite{zakrzewski}, \cite{kowclas})  one can distinguish three kinds
of quantum
spaces:\\
\\
{ 1)} Canonical ($\theta^{\mu\nu}$-deformed) space-time
\begin{equation}
[\;{\hat x}_{\mu},{\hat x}_{\nu}\;] = i\theta_{\mu\nu}\;\;\;;\;\;\;
\theta_{\mu\nu} = {\rm const}\;, \label{noncomm}
\end{equation}
considered in  \cite{3e}-\cite{3c}. The corresponding twist
deformation of  Poincar\'{e} Hopf algebra
$\mathcal{U}_\theta(\mathcal{P})$ has been proposed in \cite{3a},
while its dual quantum group $\mathcal{P}_{\theta}$  in \cite{3e}
and \cite{3c}. There were also provided   two
$\theta^{\mu\nu}$-deformed Galilei Hopf algebras \cite{dasz} as the
contraction limits of twisted  Poincar\'{e} group $\mathcal{U}_\theta(\mathcal{P})$. \\
\\
{ 2)} Lie-algebraic modification of classical space
\begin{equation}
[\;{\hat x}_{\mu},{\hat x}_{\nu}\;] = i\theta_{\mu\nu}^{\rho}{\hat
x}_{\rho}\;, \label{noncomm1}
\end{equation}
with  particularly chosen coefficients $\theta_{\mu\nu}^{\rho}$
being constants. There exist two explicit realizations of such a
noncommutativity - $\kappa$-Poincar\'{e} Hopf algebra
$\mathcal{U}_\kappa(\mathcal{P})$  \cite{4a}, \cite{4b} and  twisted
Poincar\'{e}  group $\mathcal{U}_\zeta(\mathcal{P})$ \cite{lie2}
(see also \cite{lie1}). Their dual partners $\mathcal{P}_\kappa$ and
$\mathcal{P}_\zeta$ have been recovered in \cite{zakrzewskigrupa}
and \cite{lie2}, respectively. Besides, the so-called
$\kappa$-Galilei
 group has been provided by nonrelativistic
contraction of $\kappa$-Poincar\'{e}  Hopf algebra in
\cite{kappaga}, and its dual quantum partner has been described in
\cite{gg}. The remaining Galilei  algebras were recovered in
\cite{dasz} by various contractions   of
twisted Poincar\'{e}  group $\mathcal{U}_\zeta(\mathcal{P})$.\\
\\
{ 3)} Quadratic deformation of Minkowski space
\begin{equation}
[\;{\hat x}_{\mu},{\hat x}_{\nu}\;] =
i\theta_{\mu\nu}^{\rho\tau}{\hat x}_{\rho}{\hat x}_{\tau}\;,
\label{noncomm2}
\end{equation}
with coefficients $\theta_{\mu\nu}^{\rho\tau}$ being constants. This
type of noncommutativity has been proposed   as the translation
sector of   Poisson-Lie structure $\mathcal{P}_\xi$. The explicit
form of its
nonrelativistic counterpart remains unknown. \\

In this article we perform three nonrelativistic contractions (see
Section 2) of  twisted Poincar\'{e} groups $\mathcal{P}_\theta$,
$\mathcal{P}_\zeta$ and $\mathcal{P}_\xi$, respectively. In such a
way we recover six Galilei Hopf algebras  dual to the quantum
(Galilei) groups proposed in  \cite{dasz}\footnote{They were
obtained by contractions of $\mathcal{U}_\theta(\mathcal{P})$,
$\mathcal{U}_\zeta(\mathcal{P})$ and $\mathcal{U}_\xi(\mathcal{P})$
twisted Poincar\'{e} Hopf algebras.}. Two of them correspond to
canonical (1)), three - to Lie-algebraic (2)), and one - to
quadratic (3)) type of space-time
noncommutativity. We show that in the Lie-algebraic 
case their translation sectors can be  identified with the
nonrelativistic space-times, introduced in \cite{dasz} as a quantum
representation space (a Hopf module) of twisted (Galilei) algebras.
Consequently, we reproduce three Lie-algebraically deformed
space-times: two  with quantum space and classical time, and one
with classical space and quantum time. In such a way  we also
recover a new quadratic Galilei space-time (a translation sector)
with quantum space and classical time.

It should be mentioned that presented  groups can be recovered with
use of two other Hopf-algebraic methods \cite{frt}, \cite{poisson}.
First of them, so-called FRT procedure \cite{frt}, uses quantum
R-matrix associated with the considered algebra, while the second
one, leads to  quantum group by canonical quantization of a suitable
Poisson-Lie
structure \cite{poisson}, \cite{kowclas}. 
It should be noted, however, that contraction scheme used in this
article  has one advantage - it gives an additional information
about relativistic counterparts of recovered algebras, i.e. we get
our Galilei groups as a contraction limit of existing Poincar\'{e}
Hopf structures.

The knowledge of explicit form  of Galilei   Hopf algebra
$(\mathcal{U}_{.}(\mathcal{G}))$ as well as its dual quantum group
$(\mathcal{G}_{.})$ permits us to analyze the basic nonrelativistic
dynamical systems. Using so-called Heisenberg double procedure
\cite{majid1} one can provide a proper
phase-space associated with the considered Hopf algebras. 
Such a construction in the case of relativistic symmetries has been
presented  in \cite{{phamelia}}, \cite{kappaph2} for
$\kappa$-Poincar\'{e} algebra, and in \cite{lieph} for
Lie-algebraically twisted Poincar\'{e} group. Moreover, 
the Heisenberg uncertainty principle corresponding to the above
(quantum) phase-spaces, has been provided in \cite{phamelia} and
\cite{lieph} for $\kappa$- and twist-deformed symmetries
respectively. 
The analogous investigations  at nonrelativistic level already has
been undertaken. 

The paper is organized as follows. In second Section we describe
three contraction procedures used in this article - one
$c$-independent and two with $c$-dependent  parameter of
deformation. In Sections 3, 4 and 5 we find canonical, Lie-algebraic
and quadratic twist deformations of quantum Galilei group,
respectively. The results are summarized and discussed in the last
Section.

\section{{{Contraction procedures.}}}

Let us consider the following redefinition of rotation and
translation generators of quantum Poincar\'{e} group \cite{inonu}
(see also \cite{gg})\footnote{The light-velocity $c$ plays a role of
contraction parameter.}
\begin{eqnarray}
&& {\Lambda} _{\ 0 }^{0 } = \left(
1+\frac{\overline{v}^2}{c^2}\right)^{\frac{1}{2}}\;,
\label{zadrugizm1}\\
&~~&  \cr && {\Lambda} _{\ 0 }^{i } = \frac{v^i}{c}\;,\\
&~~&  \cr
&& {\Lambda} _{\ i }^{0 }= \frac{v^k{R} _{\ i}^{k }}{c}\;,\\
&~~&  \cr && {\Lambda} _{\ i }^{k } = \left(\delta _{\ l }^{k
}+\left(\left(
1+\frac{\overline{v}^2}{c^2}\right)^{\frac{1}{2}}-1\right)
\frac{v^kv^l}{\overline{v}^2}\right){R} _{\ i }^{l } \;,\\
&~~&  \cr &&a^i = b^i\;\;\;,\;\;\;a^0 = c\tau\;, \label{zadrugizm7}
\end{eqnarray}
where  ${R} _{\ j }^{i },v^i,\tau,b^i$ denote the generators of
Galilei quantum group - rotations, boosts and translations,
respectively. In this article we consider three nonrelativistic
contractions:\\ 
\\
$~~$${i)}$ Standard (In\"{o}n\"{u}-Wigner) contraction with
$c$-independent parameter of deformation

\cite{inonu}, i.e. we use the redefinition (\ref{zadrugizm1})-(\ref{zadrugizm7}) and
take the $c \to \infty$ limit.\\
\\
$~$${ii)}$ Contraction with $c$-dependent parameter $\kappa$
$(\left[\,{\kappa}\,\right]=(lenght)^{-1})$ such that
${\kappa}=\hat{\kappa}/c$

$(\left[\,\hat{\kappa}\,\right]=(time)^{-1})$ \cite{azca1},
\cite{azca2}. Then, in the $c \to \infty$ limit we get the quantum
group with

deformation parameter $\hat{\kappa}$.\\
\\
${iii)}$ The parameter $\kappa$ is replaced by
${\kappa}=\overline{\kappa}c$
$(\left[\,\overline{\kappa}\,\right]=(time)\times (lenght)^{-2})$
(see e.g. \cite{azca1}),

and the contraction limit  $c \to \infty$ leads to the quantum group
with parameter $\overline{\kappa}$.
\\

As we mentioned in Introduction, we shall perform  three
contractions in the case of canonical ${\mathcal P}_\theta$,
Lie-algebraic ${\mathcal P}_{\zeta}$ and quadratic ${\mathcal
P}_\xi$ Poincar\'{e} groups. They were provided in \cite{3a},
\cite{lie2} together with their dual partners $\mathcal{U}({\mathcal
P}_\theta)$, $\mathcal{U}({\mathcal P}_\zeta)$ and
$\mathcal{U}({\mathcal P}_\xi)$, respectively.

\section{{{Canonical deformation.}}}
Let us start with canonical deformation of relativistic symmetries.
The $\theta^{\mu\nu}$-deformed Poincar\'{e} group ${\mathcal
P}_\theta$ has been proposed some years ago in \cite{3e} and
rediscovered recently in \cite{3c}. Its algebraic sector looks as
follows\footnote{$\eta_{\mu\nu}= (-,+,+,+)$.}
\begin{eqnarray}
&&[\; {a}^{\mu },{a}^{\nu }\;] =i\,\theta ^{\rho \sigma }( {\Lambda}
_{\ \rho }^{\mu }\,{\Lambda} _{\ \sigma }^{\nu }-\delta _{\ \rho
}^{\mu }\,\delta
_{\ \sigma }^{\nu }) \;,  \label{dlww4}\\
&~~&  \cr
 &&[\; {\Lambda} _{\ \tau }^{\mu },{\Lambda} _{\ \rho }^{\nu
}\;] =[\; {a}^{\mu },{\Lambda} _{\ \rho }^{\nu }\;] =0\;,
\label{dlws5}
\end{eqnarray}
while  coproducts remain undeformed
\begin{equation}
\Delta ({a}^{\mu })={\Lambda} _{\ \nu }^{\mu }\otimes {a}^{\nu
}+{a}^{\mu }\otimes 1\,, \qquad \Delta (\,{\Lambda} _{\ \nu }^{\mu
})={\Lambda} _{\ \rho }^{\mu }\otimes {\Lambda} _{\ \nu }^{\rho }\;.
\label{dlww6}
\end{equation}
The corresponding classical r-matrix has the form \cite{zakrzewski}
\begin{equation}
r_{\theta}=\frac{1}{2}\theta ^{\mu \nu }\,P_{\mu }\wedge P_{\nu }\;,
\label{stwor1}
\end{equation}
with fourmomentum generators $P_\mu$  dual to the translations
$a^\mu$. Obviously, the r-matrix (\ref{stwor1}) satisfies the
classical Yang-Baxter (CYBE) equation
\begin{equation}
[[\;r_{\theta},r_{\theta}\;]] = [\;r_{\theta12},r_{\theta13} +
r_{\theta23}\;] + [\;r_{\theta13}, r_{\theta23}\;] = 0\;,
\label{cybe}
\end{equation}
where    symbol $[[\;\cdot,\cdot\;]]$ denotes the Schouten bracket
and $r_{\theta 12} = \frac{1}{2}\theta ^{\mu \nu }P_{\mu }\wedge
P_\nu\wedge 1$, $r_{\theta 13} = \frac{1}{2}\theta ^{\mu \nu }P_{\mu
}\wedge 1 \wedge P_\nu$, $r_{\theta 23} = \frac{1}{2}\theta ^{\mu
\nu }1\wedge P_\mu\wedge P_\nu$.

In the case of simplest contraction  of ${\mathcal P}_\theta$ group
(see ${i)}$), for parameter $c$ running to infinity, we get the
following algebraic sector
\begin{eqnarray}
&&~~~~[\; {b}^{i },{b}^{j }\;] =i\,\theta ^{kl }( {R} _{\ k}^{i
}\,{R} _{\ l }^{j }-\delta _{\ k }^{i }\,\delta
_{\ l }^{j }) \;,  \label{x1}\\
&~~&  \cr &&~~~~[\; {{\tau}},b^{i }\;]=[\; {{\tau}},v^{i }\;] = [\;
b^i,v^{j }\;]= [\; v^i,v^{j }\;]=0\;, \\&~~& \cr &&[\; {R} _{\ j
}^{i },{R} _{\ l }^{k}\;]=[\; {v}^i,{R} _{\ l }^{k }\;] =[\;
{{\tau}},{R} _{\ j }^{i }\;] = [\; b^i,{R} _{\ l }^{k }\;] =0\;,
\label{x2}
\end{eqnarray}
supplemented by the classical (undeformed) coproducts
\begin{eqnarray}
&&\Delta ({R}_{\ j }^{i}) = {R}_{\ k }^{i }\otimes {R}_{\
j}^{k }\;, \label{toporzelx}\\
&~~&  \cr &&\Delta ({v}^{i })={R}_{\ j}^{i}\otimes {v}^{j }+{v}^{i
}\otimes 1\;,\\
&~~&  \cr &&\Delta (\tau)= \tau\otimes 1+1\otimes \tau\;,\\ &~~& \cr
 &&\Delta ({b}^{i })={R}_{\ j}^{i }\otimes {b}^{j
}+{v}^{i }\otimes \tau+{b}^{i }\otimes 1\;. \label{toporzelx1}
\end{eqnarray}
The relations (\ref{x1})-(\ref{toporzelx1}) define softly deformed
Galilei group ${\mathcal G}_\theta$ dual to the  Hopf algebra
$\mathcal{U}_\theta(\mathcal{G})$ associated with the  following
classical $\texttt{r}_{\theta}$-matrix  (see
\cite{dasz})\footnote{All mentioned in this article classical
r-matrices satisfy the classical Yang-Baxter equation (\ref{cybe}),
i.e. they correspond to twist-deformed Galilei Hopf algebras.}
\begin{equation}
\texttt{r}_{\theta}=\frac{1}{2}\theta ^{kl}\,\Pi_{k}\wedge
\Pi_{l}\;\;,\;\;\;\Pi_{i} -    {\rm dual\; to\; the\;
translations}\; b^i\;.\label{grmatix}
\end{equation}

Let us  turn to the contraction ${ii)}$ of ${\mathcal P}_\theta$
group $(\theta^{\mu\nu} = {\hat{\theta}^{\mu\nu}}/{\kappa})$ with
$c$-dependent deformation parameter $\kappa = {\hat \kappa}/c$. In
such a case, for ${\hat{\theta}}^{ij}=0$ and ${\hat \theta}^{0i}=
\frac{{{\theta}^{0i}}c}{\hat\kappa}$, in the contraction limit $c
\to \infty$ we get
\begin{eqnarray}
&&~[\; {b}^{i },{b}^{j }\;] =i\,\frac{{\hat \theta} ^{0k }}{{\hat
\kappa}}( v^{i }\,{R} _{\ k }^{j
}-{R} _{\ k }^{i}v^j ) \;,  \label{mit1}\\
&~~&  \cr &&~[\; {{\tau}},b^{i }\;]=i\,\frac{{\hat \theta} ^{0k
}}{{\hat \kappa}}
({R} _{\ k }^{i}+ \delta_{\ k }^{i})\;,\\
&~~&  \cr &&[\; {{\tau}},v^{i }\;] = [\; b^i,v^{j }\;]= [\; v^i,v^{j
}\;]=0\;,
\\&~~& \cr &&[\; {R} _{\ j }^{i },{R} _{\ l }^{k}\;]=[\; {v}^i,{R}
_{\ l }^{k }\;] =[\; {{\tau}},{R} _{\ j }^{i }\;] = [\; b^i,{R} _{\
l }^{k }\;] =0\;, \label{mit2}
\end{eqnarray}
with the classical coproduct (\ref{toporzelx})-(\ref{toporzelx1})
and corresponding classical
$\texttt{r}_{\frac{\hat{\theta}}{\hat{\kappa}}}$-matrix
\begin{equation}
\texttt{r}_{\frac{\hat{\theta}}{\hat{\kappa}}}=\frac{\hat{\theta}^{0k}}{\hat{\kappa}}\,\Pi_{0}\wedge
\Pi_{k}\;\;;\;\;\;\Pi_{0} - {\rm dual\;\,to\;\,the\;\,generator}\;\,
\tau\;. \label{hhgrmatix}
\end{equation}
The relations (\ref{mit1})-(\ref{mit2}) and
(\ref{toporzelx})-(\ref{toporzelx1}) define softly deformed Galilei
 group ${\mathcal G}_{\frac{\hat\theta}{\hat\kappa}}$ dual to
the algebra
$\mathcal{U}_{\frac{\hat\theta}{\hat\kappa}}(\mathcal{G})$ (see
\cite{dasz}). One can also check that for $\theta^{ij}\ne 0$ the
contraction ${ii)}$ becomes divergent, while in the case of
contraction ${iii)}$ $(\kappa = {\overline{ \kappa}}c)$, for
arbitrary value of parameter $\theta^{\mu\nu}$, we get the
undeformed Galilei quantum group $\,{\mathcal G}_{0}$.

\section{{{Lie-algebraic deformation.}}}

The Lie-algebraic twist deformation of Poincar\'{e} group ${\mathcal
P}_{\zeta}$ has been studied in \cite{lie2}. Its algebraic sector
looks as follows
\begin{eqnarray}
&&[\; {a}^{\mu },{a}^{\nu }\;] =i\zeta ^{\nu } ( \delta _{\ \alpha
}^{\mu }{a}_{\beta }-\delta _{\ \beta }^{\mu }{a}_{\alpha }) +
i\zeta^\mu( \delta _{\ \beta }^{\nu }{a}_{\alpha }-\delta _{\ \alpha
}^{\nu
}{a}_{\beta })\;,  \label{dlww2.10a} \\
&~~&  \cr &&[\; {a}^{\mu },{\Lambda} _{\ \rho }^{\nu }\;] =i\zeta
^{\lambda }{\Lambda} _{\ \lambda }^{\mu }( \eta _{\beta \rho
}{\Lambda} _{\ \alpha }^{\nu }-\eta _{\alpha \rho }{\Lambda} _{\
\beta }^{\nu }) +i\zeta ^{\mu }( \delta _{\ \beta }^{\nu }{\Lambda}
_{\alpha \rho }-\delta _{\ \alpha }^{\nu }{\Lambda} _{\beta \rho })
\;,
\label{dlww2.10b} \\
&~~&  \cr &&~~~~~~~~~~~~~~~~~~~~~~~~~~[\; {\Lambda} _{\ \nu }^{\mu
},{\Lambda} _{\ \tau }^{\rho }\;] =0\;, \label{dlww2.10c}
\end{eqnarray}
while coproducts remain classical (see (\ref{dlww6})). The
corresponding $r_\zeta$-matrix has the form
\begin{equation}
r_\zeta= \frac{1}{2}\zeta^\lambda\,P_{\lambda }\wedge M_{\alpha
\beta }\;\;;\;\;\; \lambda \ne \alpha,\;\beta  - {\rm fixed} \;,
\label{swiatowid}
\end{equation}
where generators $M_{\mu\nu}$ are dual  to the  rotations ${\Lambda}
_{\ \nu }^{\mu }$.

In the case of "space-like" carrier $\{\,M_{kl}, P_\gamma\;;\;
\gamma \ne\,k,\,l,\,0\,\}$ the $c$-independent contraction ${i)}$
leads to the following algebraic sector
\begin{eqnarray}
&&[\; {b}^{i },{b}^{j }\;] =i\,\zeta \delta^{j\gamma}( \delta _{\
k}^{i }\,{b} _{l }-\delta _{\ l }^{i }\, b_{k }) +
i\,\zeta\delta^{i\gamma}( \delta _{\ l}^{j }\,{b} _{k }-\delta _{\ k
}^{j }\,
b_{l })\;,  \label{system}\\
&~~&  \cr &&[\; {b}^{i },{v}^{j }\;] =i\,\zeta \delta^{i\gamma}(
\delta _{\ l}^{j }\,{v} _{k }-\delta _{\ k }^{j }\,
v_{l }) \;,  \label{supersystem}\\
&~~&  \cr &&[\; b^i,{R} _{\ \tau }^{\rho}\;]=i\zeta {R} _{\
\gamma}^{i }( \delta _{l\tau}{R} _{\ k }^{\rho }-\delta _{k \tau
}{R} _{\ l }^{\rho }) +i\zeta \delta_{\ \gamma}^{i}( \delta _{\
l}^{\rho }{R} _{k\tau }-\delta _{\ k }^{\rho }{R} _{l\tau })
\;,\\
&~~&  \cr &&[\; {{\tau}},b^{i }\;]=[\; {{\tau}},v^{i }\;] =  [\;
v^i,v^{j }\;]=0\;,\label{perun} \\&~~& \cr &&[\; {R} _{\ j }^{i
},{R} _{\ l }^{k}\;]=[\; {v}^i,{R} _{\ l }^{k }\;] =[\; {{\tau}},{R}
_{\ j }^{i }\;]  =0\;, \label{system1}
\end{eqnarray}
and   undeformed coproducts (\ref{toporzelx})-(\ref{toporzelx1}).
The above relations define  twisted Galilei group $\,{\mathcal
G}_{\zeta}$ dual to the   Hopf algebra
$\mathcal{U}_{\zeta}(\mathcal{G})$ (see \cite{dasz}).  In the case
of carrier $\{\,M_{kl}, P_0 \,\}$ we get undeformed Galilei quantum
group $\,{\mathcal G}_{0}$, while for "boost-like" carrier
 $\{\,M_{k0}, P_l \;;\; k
\ne\,l\,\}$ the commutators of boosts with translations become
divergent.

 It should
be noted that  Hopf algebra $\mathcal{U}_{\zeta}(\mathcal{G})$ has
been provided with use of the following twist factor
\begin{equation}
\mathcal{K}_{\zeta}=  {\rm \exp}
\,\frac{i}{2}(\zeta\,\Pi_{\gamma}\wedge K_{kl})\;\;;\;\;\;K_{kl} -
{\rm dual\;\,to\;\,the\;\,generator}\;\, R^k_{~l} \;.
\label{swarozyc}
\end{equation}
In such a case  one can define the corresponding nonrelativistic
space-time
 as its  quantum
representation space - a Hopf module \cite{bloch}. It looks as
follows \cite{dasz}\footnote{$[{a},{b}]_{\star_{{\zeta}}}\equiv
a{\star_{{\zeta}}}b -b{\star_{{\zeta}}}a\,$.}
\begin{eqnarray}
&&[\, x_{i },x_{j }\,] _{\star_{{\zeta }} }= i\zeta \delta_{\gamma
j}(
 \delta _{k i }x_{l }- \delta_{l i }x_{k }) +i\zeta \delta_{\gamma
i}(\delta _{ l j }x_{k } - \delta _{k j }x_{l }
)\;,\label{ssstarswar0}\\&~~&  \cr
&&~~~~~~~~~~~~~~~~~~~~~~[\,t,x_i\,]_{{\star}_{{\zeta }}} = 0\;,
\label{ssstarswar1}
\end{eqnarray}
where the ${\star}_{{\zeta }}$-multiplication of two functions is
given by
\begin{equation}
f(t,\overline{x})\star_{{\zeta}} g(t,\overline{x}):=
\omega\circ\left(
 \mathcal{K}_{\zeta}^{-1}\rhd  f(t,\overline{x})\otimes g(t,\overline{x})\right)
 \;,
\label{starrr}
\end{equation}
with $\mathcal{K}_{\zeta}=  {\rm \exp}
\,\left(-\frac{i}{2}\zeta\,\partial_{\gamma}\wedge ( x_{k
}{\partial_l} -x_{l }{\partial_k})\right)$ and $\omega\circ\left(
a\otimes b\right) = a\cdot b$. Hence, we see, that after the
substitution
\begin{equation}
\tau \leftrightarrow t \;\;\;,\;\;\;b^i \leftrightarrow
x^i\label{substy}\;,
\end{equation}
the translation sector (\ref{system}), (\ref{perun}) can be
identified with the nonrelativistic quantum space-time
(\ref{ssstarswar0}), (\ref{ssstarswar1}).

Let us now turn to the contraction ${ii)}$ of the Poincare group
$\,{\mathcal P}_{\zeta}$ with ${\zeta}=\frac{1}{\kappa}$. Then, for
$\{\,M_{kl}, P_0\,\}$ the corresponding Galilei quantum group
$\,{\mathcal G}_{{\hat\kappa}}$ $(\kappa = {\hat \kappa}/c)$ looks
as follows
\begin{eqnarray}
&&[\; {{\tau}},b^{i }\;] =\frac{i}{{\hat\kappa}} ( \delta _{\ k }^{i
}\, b_{l }-\delta _{\ l}^{i
}\,{b} _{k }) \;,  \label{ssystem}\\
&~~&  \cr &&[\; {{\tau}},v^{i }\;] = \frac{i}{{\hat\kappa}} ( \delta
_{\ k }^{i }\, v_{l }-\delta
_{\ l}^{i }\,{v} _{k }) \;,\\
&~~&  \cr &&[\; {{\tau}},{R} _{\ \tau }^{\rho}\;]
=\frac{i}{{\hat\kappa}}( \delta _{l\tau}{R} _{\ k }^{\rho }-\delta
_{k \tau }{R} _{\ l }^{\rho }) + \frac{i}{{\hat\kappa}}( \delta _{\
k }^{\rho }{R} _{ l \tau }-\delta _{\
l }^{\rho }{R} _{ k \tau })\;,\\
&~~&  \cr &&[\; b^i,{R} _{\ \tau
}^{\rho}\;]=\frac{i}{{\hat\kappa}}v^{i }( \delta _{l\tau}{R} _{\ k
}^{\rho }-\delta _{k \tau }{R} _{\ l }^{\rho })
\;,\\
&~~&  \cr &&[\; {b}^{i },{b}^{j }\;]= [\; {b}^{i },{v}^{j }\;] = [\;
v^i,v^{j }\;]=0\label{nadnarodslawski}\;,
\\&~~& \cr &&[\; {R} _{\ j }^{i },{R} _{\ l }^{k}\;]=[\; {v}^i,{R}
_{\ l }^{k }\;] =0\;, \label{ssystem1}
\end{eqnarray}
while coproducts remain  classical. For carriers $\{\,M_{kl},
P_\gamma\;;\; \gamma \ne\,k,\,l,\,0\,\}$ and $\{\,M_{k0}, P_l\;;\; k
\ne\,l\,\}$ the contraction ${ii)}$ becomes divergent. Besides, it
should be  noted that a proper (dual) Hopf  algebra
$\mathcal{U}_{{\hat\kappa}}(\mathcal{G})$ and the corresponding
${\hat\kappa}$-deformed space-time\footnote{The above space-time is
equipped with quantum time and classical space. For its
$\mathcal{N}=1$ supersymmetric counterpart see \cite{susydasz}.}
\begin{equation}
[\, t,x_{i }\,]_{{\star}_{{\hat\kappa}}}= \frac{i}{{\hat\kappa}} (
\delta _{ li }x_{k }-\delta _{k i }x_{l } )\;\;\;,\;\;\;[\,x_{i
},x_j\,]_{{\star}_{{\hat\kappa}}} =  0 \;, \label{sistar}
\end{equation}
have been provided in \cite{dasz}\footnote{The corresponding twist
factor looks as follows $\mathcal{K}_{\hat {\kappa}}=  {\rm \exp}
\,(\frac{i}{2\hat {\kappa}}\Pi_0\wedge K_{kl})$
.}. We see, that after substitution (\ref{substy}) the relations
(\ref{sistar}) and (\ref{ssystem}), (\ref{nadnarodslawski}) become
identical.

Let us now consider the contraction ${iii)}$. Then, for carrier
$\{\,M_{kl}, P_\gamma\;;\; \gamma \ne\,k,\,l,\,0\,\}$ we obtain  the
classical (undeformed) Galilei Hopf algebra $\,{\mathcal G}_{0}$.

In the case $\{\,M_{k0}, P_l\;;\; k \ne\,l\,\}$ the situation is
more complcated, i.e. one can check that after contraction we get
the following Galilei quantum group $\,{\mathcal
G}_{\overline{\kappa}}$ $(\kappa = {\overline{\kappa}}c)$
\begin{eqnarray}
&& [\; {b}^{i },{b}^{j }\;]= \frac{i}{{\overline\kappa}} \tau(
\delta
_{\ l}^{j }\delta _{\ k}^{i }\,-\delta _{\ l }^{i }\delta _{\ k}^{j }) \;,\label{sssystem}\\
&~~&  \cr  &&[\; {v}^{i },{b}^{j }\;]=\frac{i}{{\overline\kappa}}(
{R} _{\ l }^{j }{R} _{\ k }^{i }+\delta _{\ l }^{j }\delta _{\ k
}^{i })
\;,\\
&~~&  \cr && [\; {{\tau}},b^{i }\;]=[\; {{\tau}},v^{i }\;]=[\;
{{\tau}},{R} _{\ j}^{i}\;]= [\; v^i,v^{j }\;]=0\;,\label{stachniuk1}
\\&~~& \cr &&[\; b^i,{R} _{\ \tau
}^{\rho}\;]=[\; {R} _{\ j }^{i },{R} _{\ \tau }^{\rho}\;]=[\;
{v}^i,{R} _{\ \tau }^{\rho}\;] =0\;, \label{sssystem1}
\end{eqnarray}
with trivial coproduct (\ref{toporzelx})-(\ref{toporzelx1}).

The corresponding (dual) Hopf algebra
$\mathcal{U}_{{\overline\kappa}}(\mathcal{G})$ has been recovered
with use of the following twist factor
\begin{equation}
{\mathcal{K}}_{\overline{\kappa}}=\exp
\frac{i}{2\overline{\kappa}}(\Pi_{l}\wedge V_k)\;\;;\;\;\;V_k - {\rm
dual\;\,to\;\,the\;\,boost\;\, generator}\;\, v_k \;.
\label{dghfactor}
\end{equation}
One can  observe that its nonrelativistic space-time is exactly the
same as the translation sector (\ref{sssystem}), (\ref{stachniuk1})
(see \cite{dasz})
\begin{equation}
[\, x_{i },x_{j }\,] _{\star_{\overline{\kappa}}}=
\frac{i}{\overline{\kappa}}t(\delta _{l i }\delta_{k j }-\delta _{
ki }\delta _{l j
})\;\;\;,\;\;\;[\,t,x_i\,]_{\star_{\overline{\kappa}}} = 0 \;,
\label{ysesstar}
\end{equation}
where $\star_{\overline{\kappa}}$-multiplication is given by
\begin{equation*}
f(t,\overline{x})\,{\star_{\overline{\kappa}}}\, g(t,\overline{x}):=
\omega\circ\left(
 \mathcal{K}_{{\overline{\kappa}}}^{-1}\rhd  f(t,\overline{x})\otimes
 g(t,\overline{x})\right)\;\;;\;\;\;\mathcal{K}_{\overline{\kappa}}=  {\rm \exp}
(\frac{i}{2\overline{\kappa}}\,\partial_{l}\wedge t{\partial_k} )
 \;,
\end{equation*}
i.e. we can identify the translation sector (\ref{sssystem}),
(\ref{stachniuk1}) with nonrelativistic space-time (\ref{ysesstar}).

\section{{{Quadratic deformation.}}}

The quadratic deformation of Poincar\'{e} group  ${\mathcal P}_\xi$
has been investigated  in \cite{lie2}. It is given by the following
algebraic sector
\begin{eqnarray}
[\;{\Lambda}^\mu_{\ \tau}, {\Lambda}^\nu_{\ \rho}\;]&=& (1-\cosh\xi)
\sum_{\genfrac{}{}{0pt}{}{k=\alpha,\beta}{l=\gamma,\delta}}^{}
(\delta^{\mu}_ {\ \{k}\delta^{\nu}_{\ l\}}{\Lambda}^{\{k}_{\ \tau}
{\Lambda}^{l\}}_{\ \rho}
-\delta^{\{k}_ {\ \rho}\delta^{l\}}_{\ \tau}{\Lambda}^\nu_{\ \{k} {\Lambda}^\mu_{\ l\}})+\label{qrtt}\\
&~~&  \cr &+&i\sinh\xi[(\eta_{\beta \rho}{\Lambda}^\nu_{\
\alpha}-\eta_{\alpha \rho}{\Lambda}^\nu_{\ \beta})
(\eta_{\delta \tau}{\Lambda}^\mu_{\ \gamma}-\eta_{\gamma \tau}{\Lambda}^\mu_{\ \delta})+ \nonumber\\
&~~&  \cr &+&(\eta_{\delta \rho}{\Lambda}^\nu_{\
\gamma}-\eta_{\gamma \rho}{\Lambda}^\nu_{\ \delta})
(\eta_{\alpha \tau}{\Lambda}^\mu_{\ \beta}-\eta_{\beta \tau}{\Lambda}^\mu_{\ \alpha})+\nonumber\\
&~~&  \cr &+&(\eta^{\beta \mu}{\Lambda}^\alpha_{\ \tau}-\eta^{\alpha
\mu}{\Lambda}^\beta_{\ \tau})
(\eta^{\gamma \nu}{\Lambda}^\delta_{\ \rho}-\eta^{\delta \nu}{\Lambda}^\gamma_{\ \rho})+\nonumber\\
&~~&  \cr &+&(\eta^{\gamma \mu}{\Lambda}^\delta_{\
\tau}-\eta^{\delta \mu}{\Lambda}^\gamma_{\ \tau}) (\eta^{\alpha
\nu}{\Lambda}^\beta_{\ \rho}-\eta^{\beta \nu}{\Lambda}^\alpha_{\
\rho})]\;,\nonumber
\end{eqnarray}
\begin{eqnarray}
 ~~~~~~  [\;{a}^\mu,{a}^\nu\;]&=&\frac{i}{2}\sinh\xi(
\delta^{[\mu}_{\ \alpha}\delta^{\nu]}_{\
\gamma}\;\{{a}_\beta,{a}_\delta\}
-\delta^{[\mu}_{\ \alpha}\delta^{\nu]}_{\ \delta}\;\{{a}_{\beta},{a}_\gamma\}+~~~~~~~~~~~~~~~~~~~~~\\
&~~&  \cr &-&\delta^{[\mu}_{\ \beta}\delta^{\nu]}_{\
\gamma}\;\{{a}_\alpha,{a}_\delta\}
+\delta^{[\mu}_{\ \beta}\delta^{\nu]}_{\ \delta}\;\{{a}_\alpha,{a}_\gamma\})+\nonumber\\
&~~&  \cr &+&\frac{1}{2}(1-\cosh
\xi)\sum_{\genfrac{}{}{0pt}{}{k=\alpha,\beta}{l=\gamma,\delta}}^{}
\delta_{\ k}^{[\mu}\delta_{\
l}^{\nu]}\;[\;{a}^k,{a}^l\;]\;,\nonumber
\end{eqnarray}
\begin{eqnarray}
[\;{a}^\mu, {\Lambda}^\nu_{\ \rho}\;]&=&(1-\cosh\xi)
\sum_{\genfrac{}{}{0pt}{}{k=\alpha,\beta}{l=\gamma,\delta}}^{}
\delta^{\mu}_ {\ \{k}\delta^{\nu}_{\ l\}}{a}^{\{k} {\Lambda}_{\ \rho}^{l\}}+\\
    &+&i\sinh\xi[\;(\delta^{\mu}_{\ \alpha}{a}_\beta-\delta^{\mu}_{\ \beta}{a}_\alpha)
(\delta^{\nu}_{\ \gamma}{\Lambda}_{\delta \rho}-\delta^{\nu}_{\ \delta}{\Lambda}_{\gamma \rho})+\nonumber\\
&~~&  \cr &-&(\delta^{\mu}_{\ \gamma}{a}_\delta-\delta^{\mu}_{\
\delta}{a}_\gamma) (\delta^{\nu}_{\ \alpha}{\Lambda}_{\beta
\rho}-\delta^{\nu}_{\ \beta}{\Lambda}_{\alpha
\rho})\;]\;,~~~~~~~~~~~~~~~~~~~~~~\nonumber
\end{eqnarray}
~\\
with $\delta^{\mu}_ {\ \{k}\delta^{\nu}_{\
l\}}{\mathcal{O}}^{\{k}_{\ \tau} {\mathcal{O}}^{l\}}_{\ \rho}
=\delta^{\mu}_ {\ k}\delta^{\nu}_{\ l}{\mathcal{O}}^{k}_{\ \tau}
{\mathcal{O}}^{l}_{\ \rho}+ \delta^{\mu}_ {\ l}\delta^{\nu}_{\
k}{\mathcal{O}}^{l}_{\ \tau} {\mathcal{O}}^{k}_{\ \rho}$, and the
classical coproduct (\ref{dlww6}). The corresponding $r_\xi$-matrix
looks as follows
\begin{equation}
r_\xi=\frac{1}{2}\xi M_{\alpha\beta}\wedge M_{\gamma\delta}\;,
\end{equation}
where   indices $\alpha,\,\beta\,, \gamma\,,\delta$ are all
different and fixed.

One can see that for $\alpha=i$, $\beta=0$, $\gamma=k$ and
$\delta=l$ the contraction ${i)}$ becomes divergent in  the  $c \to
\infty$ limit. Similarly, for ${\mathcal P}_\xi$ with $\xi={{\hat
\xi}}/{\kappa}$ the contraction ${ii)}$ $(\kappa={\hat{\kappa}/c})$
does not exist. In the case ${iii)}$ $(\kappa={\overline{\kappa}c})$
situation appears  less trivial, i.e. in the $c \to \infty$ limit we
get the following algebraic sector
\begin{eqnarray}
&&[\;{v}^\rho, v^{ \tau}\;] = -\frac{i{\hat
\xi}}{2\overline{\kappa}} \delta^{i\rho}(\delta^{k\tau}v^l
-\delta^{l\tau}v^k ) +\frac{i{\hat \xi}}{2\overline{\kappa}}
\delta^{i\tau}(\delta^{k\rho}v^l
-\delta^{l\rho}v^k)\;,\label{topppoorzel1}\\
&~~&  \cr
 &&[\;{R}^\rho_{\ \tau}, v^j\;] = -\frac{i{\hat
\xi}}{2\overline{\kappa}}{R}^j_{\ i}(\delta_{l\tau}{R}^\rho_{\ k}-
\delta_{k\tau}{R}^\rho_{\ l} )+ \frac{i{\hat
\xi}}{2\overline{\kappa}}\delta^i_{\ j}(\delta^{k\rho}{R}^l_{\
\tau}-\delta^{l\rho}
{R}^k_{\ \tau} )\;,\\
&~~&  \cr
  && [\;{b}^\rho,{b}^\tau\;]=\frac{i{\hat \xi}}{2\overline{\kappa}}(
\delta^{[\rho}_{\ i}\delta^{\tau]}_{\ k}\;\{\tau,{b}_l\}
-\delta^{[\rho}_{\ i}\delta^{\tau]}_{\
l}\;\{\tau,{b}_k\})\;\;\;,\;\;\;[\;\tau,{b}^j\;] =
0\;,\label{trans1}\\
&~~&  \cr &&[\;{b}^j,{R}^\rho_{\ \tau}\;]=\frac{i{\hat
\xi}}{2\overline{\kappa}}\delta^{j}_{\ i}\tau(\delta^{\rho}_{\
k}{R}_{l \tau}-\delta^{\rho}_{\ l}{R}_{k \tau})\;,\\
&~~&  \cr &&[\;\tau,{R}^\rho_{\ \tau}\;]= [\;\tau,v^j\;] =[\;{R}^\rho_{\ \tau},{R}^\sigma_{\ \omega}\;]= 0\;,\\
&~~&  \cr &&[\;b^\rho, v^{\tau}\;]=\frac{i{\hat
\xi}}{2\overline{\kappa}}\delta^{\rho}_{\ i}\tau(\delta^{\tau}_{\
k}{v}_{l}-\delta^{\tau}_{\ l}{v}_{k }) - \frac{i{\hat
\xi}}{2\overline{\kappa}}\delta^{\tau}_{\ i}(\delta^{\rho}_{\
k}{b}_{l}-\delta^{\rho}_{\ l}{b}_{k })\;,\label{topppoorzel2}
\end{eqnarray}
and classical coproducts. The above relations define quadratic
Galilei  group ${\mathcal G}_{\overline{\kappa}}$ equipped with the
following   classical $\texttt{r}_{\overline{\kappa}}$-matrix
\begin{eqnarray}
\texttt{r}_{\overline{\kappa}} =\frac{{\hat
\xi}}{2\overline{\kappa}}(V_i\wedge K_{kl} )\;. \label{toporzel}
\end{eqnarray}

Let us now turn to nonrelativistic  space-time corresponding to the
 quantum group (\ref{topppoorzel1})-(\ref{topppoorzel2}). As it was mentioned, it  is defined as a
quantum representation space  of Galilei group ${\mathcal
G}_{\overline{\kappa}}$. The action of dual generators $V_i$ and
$K_{kl}$  on such a space is given by

\begin{equation}
K_{kl}\rhd f(t,\overline{x}) =i\left( x_{k }{\partial_l} -x_{l
}{\partial_k} \right) f(t,\overline{x})\;\;\;,\;\;\; V_i\rhd
f(t,\overline{x}) =it{\partial_i} \,f(t,\overline{x})\;,\label{dsf}
\end{equation}
\\
while the $\star_{\overline{\kappa}}$-multiplication   looks as
follows
\begin{equation*}
f(t,\overline{x})\star_{{\overline{\kappa}}} g(t,\overline{x}):=
\omega\circ\left(
 \mathcal{K}_{{\overline{\kappa}}}^{-1}\rhd  f(t,\overline{x})\otimes g(t,\overline{x})\right)
 \;\;;\;\;\;\mathcal{K}_{{\overline{\kappa}}} =   {\rm \exp}
\,(\frac{i{\hat \xi}}{2{\overline{\kappa}}}\,t\partial_{i}\wedge (
x_{k }{\partial_l} -x_{l }{\partial_k}))\;.
\end{equation*}
Hence, we have

\begin{equation}
[\;{x}^\rho,{x}^\tau\;]_{\star_{{\overline{\kappa}}}}=\frac{i{\hat
\xi}}{2\overline{\kappa}}( \delta^{[\rho}_{\ i}\delta^{\tau]}_{\
k}\;\{t,{x}_l\}_{\star_{{\overline{\kappa}}}} -\delta^{[\rho}_{\
i}\delta^{\tau]}_{\
l}\;\{t,{x}_k\}_{\star_{{\overline{\kappa}}}})\;\;\;,\;\;\;
[\;t,{x}^j\;]_{\star_{{\overline{\kappa}}}} = 0
 \;.
\label{tworzycielskaslawia}
\end{equation}
\\
We see that the noncommutative space-time
(\ref{tworzycielskaslawia}) is exactly the same as the translation
sector (\ref{trans1}) attached to  quantum space and classical time.

\section{{{Final remarks.}}}

In this article we propose  canonical and Lie-algebraic twist
deformations of quantum groups dual to the recovered in \cite{dasz}
 Galilei Hopf algebras $\mathcal{U}_{.}(\mathcal{G})$. Besides, we also
 obtained
 a new quadratic deformation of
Galilei Hopf structure as well as the corresponding noncommutative
(quantum) space-time. In such a way we show that the translation
sectors of dual groups are identical with Hopf-modules (space-times)
of corresponding Galilei Hopf
algebras \cite{dasz}. 

It should be mentioned that presented results complete our studies
 in \cite{dasz} on the contractions of twisted Poincar\'{e}
groups. Nevertheless, they  can be extended in various ways. First
of all, one can consider basic dynamical models corresponding to
considered Galilei algebras  \cite{proclw}. Besides, it seems
interesting to find other $D=3+1$ dimensional nonrelativistic
space-times, corresponding Hopf algebras describing symmetry, and
dual  groups, predicted by the general classification of all
Galilean Poisson-Lie structures \cite{kowclas} (see also for
two-dimensional case \cite{lodz1} and \cite{lodz2}). One can also
ask about a "superposition" of discussed quantum deformations  as
well as their supersymmetric $\mathcal{N}=1$ extensions (see e.g.
\cite{jap} and \cite{susydasz}). Finally, as it was mentioned in
Introduction, the corresponding quantum phase-spaces can be
generated with use of the Heisenberg double procedure \cite{majid1}.
The investigation of these problems are under study.

\section*{Acknowledgments}
The author would like to thank  Jerzy Lukierski, Andrzej Frydryszak,
Marek Mozrzymas and Mariusz Woronowicz for many valuable
discussions.\\
This paper has been financially supported by Ministry of Science and
Higher Education grant NN202318534.

\end{document}